\documentclass{article}
\usepackage{spconfa4,amsmath,graphicx,tikz}
\usetikzlibrary{positioning,calc,decorations.pathreplacing}
\usetikzlibrary{shapes.geometric}

\usepackage[dvipsnames]{xcolor}

\usepackage{algorithm, algorithmic}
\usepackage{cite}
\usepackage{acronym}
\usepackage{booktabs} 
\usepackage{adjustbox} 
\usepackage{amssymb}  
\usepackage{makecell}
\usepackage{diagbox}
\usepackage{siunitx}

\usepackage{multirow} 
\usepackage{enumitem}
\usepackage{bm}  
\usepackage{pbalance}
\usepackage{graphicx}
\usepackage{tikz}
\usepackage{tikz-cd}
\usepackage{multicol}
\usetikzlibrary{arrows.meta,calc,positioning,shapes.misc}

\usepackage{pbalance}

\acrodef{SE}{speech enhancement}
\acrodef{DNN}{deep neural network}
\acrodef{SNN}{slimmable neural network}
\acrodef{DDPM}{denoising diffusion probabilistic model}
\acrodef{SDM}{slimmable diffusion model}
\acrodef{SDE}{stochastic differential equation}
\acrodef{DM}{diffusion model}
\acrodef{GAN}{generative adversarial network}
\acrodef{HC}{high-complexity}
\acrodef{MC}{medium-complexity}
\acrodef{LC}{low-complexity}

\acrodef{GAN}{generative adversarial network}
\acrodef{DoA}{direction of arrival}
\acrodef{PDF}{probability density function}
\acrodef{E2E}{end-to-end}
\acrodef{STFT}{short-time Fourier transform }
\acrodef{TF}{time-frequency}
\acrodef{RIR}{room impulse response}
\acrodef{FiLM}{feature-wise linear modulation }
\acrodef{SNR}{signal-to-noise ratio}
\acrodef{MOS}{mean opinion score}
\acrodef{FiLM}{feature-wise linear modulation }
\acrodef{TSE}{target speaker extraction }
\acrodef{SE}{speech enhancement}
\acrodef{SOTA}{state-of-the-art}
\acrodef{SDE}{stochastic differential equation}
\acrodef{ODE}{ordinary differential equation}
\acrodef{WER}{word error rate}
\acrodef{CER}{character error rate}
\acrodef{CFM}{conditional flow matching}
\acrodef{PESQ}{perceptual evaluation of speech quality}
\acrodef{ESTOI}
{extended short-time objective intelligibility }
\acrodef{SI-SIR}{scale-invariant signal-to-interference ratio}
\acrodef{FLOPs}{floating point operations}
\title{SlimDiffuSE: Towards Efficient Diffusion-Based Speech Enhancement using Slimmable Networks}
\name{
\shortstack{
Nagashree K. S. Rao$^{1}$, Shrishti Saha Shetu$^{2}$, Mohamed Elminshawi$^{1}$ \\
\textit{Emanu\"{e}l A. P. Habets}$^{1,2}$, \textit{Andreas Brendel}$^{1}$\thanks{*Andreas Brendel has been supported by the Free State of Bavaria by the DSgenAI project.}
}
}

\address{
$^{1}$Fraunhofer IIS, Am Wolfsmantel 33, 91058 Erlangen, Germany\\
$^{2}$International Audio Laboratories Erlangen\footnotemark[1], Am Wolfsmantel 33, 91058 Erlangen, Germany
\\
{\small \{nagashree.kuruduganahalli.sudhindra.rao, shrishti.saha.shetu, mohamed.elminshawi, emanuel.habets, andreas.brendel\}@iis.fraunhofer.de}
}
\begin{document}
%

\maketitle

\footnotetext[2]{A joint institution of Friedrich-Alexander-Universit\"{a}t Erlangen-N\"{u}rnberg
(FAU) and
Fraunhofer
IIS.}
\begin{abstract}
Diffusion-based models are emerging in the speech enhancement domain and are achieving state-of-the-art performance across various benchmark datasets. A major downside of diffusion models is that data generation requires many evaluations of a typically large neural network, which results in high overall complexity. In this work, we propose a slimmable diffusion model that employs adaptive network widths throughout the data generation process to reduce computational cost. By using a greedy search algorithm to optimize the network width schedule, our method achieves performance comparable to baseline diffusion models with significantly reduced computational complexity. Notably, our approach reduces the computational complexity by up to $87.5\%$ without a significant drop in objective metrics, such as \ac{PESQ} and SI-SDR.

\end{abstract}
\section{Introduction}
\label{sec:intro}
\Ac{SE} aims to improve the quality and intelligibility of speech in noisy and reverberant scenarios~\cite{SE_SBOll,cohen2003noise,cohen2001speech}. Significant progress has been achieved in recent years in \ac{SE} using \acp{DNN}~\cite{hu2020dccrn,shetu2023ultra,schroter2022deepfilternet2,DBLP:journals/corr/abs-2102-03207,chen2022fullsubnet+,shetu2024hybrid}. Most \ac{DNN}-based \ac{SE} techniques follow discriminative training paradigms, which perform well in the moderate-to-high \ac{SNR} regime; however, they exhibit a drop in performance in low \ac{SNR} scenarios~\cite{shetu2024comparative}. For such challenging scenarios, generative models gained popularity lately~\cite{shetu2025gan,fu2021metricgan+,pascual2017segan,huang2022fastdiff,richter2023speech,leglaive2020recurrent}, as they promise to reconstruct speech content that might be completely masked by noises or distortions.

Recently, \acp{DM} received a lot of attention as a promising approach to generative \ac{SE}~\cite{richter2023speech, welker2022speech,lemercier2023analysing, lemercier2023storm}. \acp{DM} are typically based on a forward noising process, which turns clean speech data into noise by adding Gaussian noise iteratively. 
This process is reversed in the backward path by denoising the signal step by step at different noise scales. \Acp{DM} can be formulated in continuous time with \acp{SDE}~\cite{song2021scorebasedgenerativemodelingstochastic} or in discrete time as \acp{DDPM}~\cite{ho2020denoising}. Despite their promising performance, \acp{DM} are computationally expensive because multiple reverse steps require repeated evaluation of large \acp{DNN}, regardless of the underlying \ac{SE} task complexity, as in SGMSE+~\cite{richter2023speech}. This makes it challenging to deploy them in real-time applications.
 

Practical \ac{SE} applications require \acp{DNN} capable of balancing performance and restrictions imposed by hardware and the application, such as memory, latency, etc. 
Several \ac{DNN} complexity reduction techniques have been explored for \ac{SE}:
A classical technique is pruning~\cite{Cheng2018ModelCA}, which eliminates redundant parameters, resulting in more lightweight models.
\ac{DNN} architectures can also be adapted during inference by varying their depth, i.e., by using a subset of layers (e.g., early exiting~\cite{early_exit}), or by varying their width, i.e., by using a subset of channels (\acp{SNN})~\cite{slimtasnet}. \Acp{SNN} use different subsets of their computational graph to avoid redundant computations during inference, which allows for
adjusting computational complexity according to the task at hand. Pruning, on the other hand, would have to maintain either multiple networks or would have to rely on a single DNN
for all considered tasks, no matter if they require a high or low computational complexity. 

In this work, based on SGMSE+, we show that different \ac{DNN} complexities are required across diffusion steps for \ac{SE}. Furthermore, we demonstrate that choosing the appropriate \ac{DNN} complexity for each diffusion step can significantly reduce overall computational cost without degrading signal quality \cite{yang2023denoising}.
Based on these findings, we propose SlimDiffuSE, a \ac{SDM} based on SGMSE+ with adaptive network width, which reduces overall computational cost while keeping the same parameter count and achieving comparable \ac{SE} performance to the original SGMSE+. 

\section{Proposed Method}
\label{sec:format}

In the SGMSE+ framework ~\cite{richter2023speech}, \ac{SE} is formulated as a conditional diffusion process in the complex short-time Fourier transform (STFT) domain. Let $\mathbf{x}_0, \mathbf{y} \in \mathbb{C}^{F \times K}$ denote a clean speech representation and the corresponding noisy observation sampled from the distributions $p_{\text{clean}}$ and $p_{\text{noisy}}$, respectively. Here, $F$ and $K$ represent the number of frequency bins and time frames, respectively.  The forward diffusion process is defined by the stochastic differential equation (SDE)
\vspace{-0.25em}
\begin{equation}
    \mathrm{d}\mathbf{x}_t = \gamma(\mathbf{y} - \mathbf{x}_t)\,\mathrm{d}t + g(t)\, \mathrm{d}\mathbf{w}_t,
\end{equation}
with drift coefficient $\gamma > 0$, time-dependent diffusion coefficient $g(t) > 0$ and standard complex-valued Wiener process $\mathbf{w}_t$. This process gradually perturbs $\mathbf{x}_0$ toward $\mathbf{y}$ by adding (Gaussian) diffusion noise scaled by $g(t)$, where  $\mathbf{x}_t \in \mathbb{C}^{F \times K}$ is the noisy signal at diffusion time $t$. The corresponding reverse-time SDE used for inference is given by
\begin{equation}
    \mathrm{d}\mathbf{x}_t = -\gamma(\mathbf{y} - \mathbf{x}_t)\,\mathrm{d}t + g^2(t)\,\mathbf{S}_\theta(\mathbf{x}_t,\mathbf{y},t)\,\mathrm{d}t + g(t)\, \mathrm{d}\bar{\mathbf{w}}_t,
\end{equation}
where $\bar{\mathbf{w}}_t$ denotes a reverse-time Wiener process and \newline $\mathbf{S}_\theta(\mathbf{x}_t,\mathbf{y},t)$ is a \ac{DNN} parameterized by $\theta$ that estimates the conditional score $\nabla_{\mathbf{x}_t} \log p_t(\mathbf{x}_t \mid \mathbf{y})$. In practice, inference is performed using a Predictor--Corrector (PC) sampler with $N$ discretization steps \cite{song2021scorebasedgenerativemodelingstochastic}, resulting in the following computational cost 
\begin{equation}
    \mathcal{C}_{\text{total}} = 2N \cdot \text{FLOPs}(\mathbf{S}_\theta),
\end{equation}
where the factor $2$ accounts for the predictor and corrector updates at each diffusion step, and $\text{FLOPs}(\mathbf{S}_\theta)$ quantifies the computational complexity of the score model in terms of \ac{FLOPs}. While effective, this formulation allocates a fixed network capacity at every diffusion step, regardless of the actual difficulty of the denoising task, which results in inefficient computational allocation across all reverse-diffusion steps. We hypothesize that during the reverse diffusion process, early time steps ($t$ close to $1$) require score models with high model capacity, as signal structure has to be built up from noise, whereas later time steps ($t$ close to $0$) primarily refine fine-grained spectral details and are comparatively less demanding (see  Sec.~\ref{ssub:models at diff compl}). Hence, assigning lower model complexities to later diffusion time steps should be sufficient for high-quality data generation. 

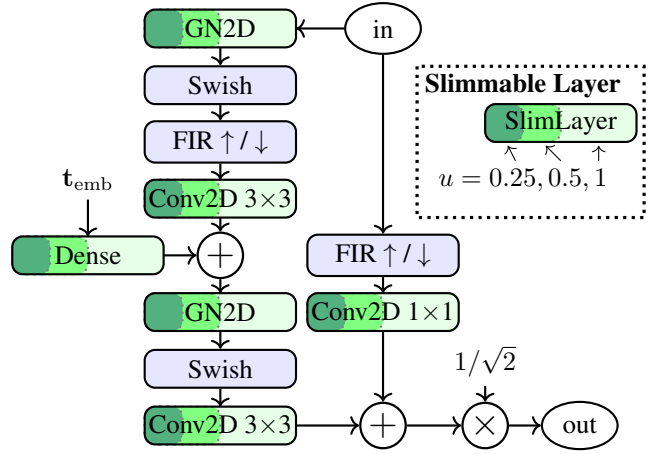
\begin{figure}[t]
\begin{tikzpicture}
\tikzset{block/.style={
    draw,
    rounded corners,
    thick,
    minimum height=0.5cm,
    minimum width=2cm,
    align=center,
    fill=blue!10
  },
  plus/.style={circle, draw, minimum size=6mm, inner sep=0pt, font=\Large, line width=0.85pt},}
\tikzset{
  pics/slimblock/.style args={#1}{
    code={
      \draw[rounded corners, fill=green!10] (-1,-0.25) rectangle (1,0.25);
      \draw[dotted, thick, gray, rounded corners, fill=green!50]
           (-1,-0.25) rectangle (0,0.25);
      \draw[dotted, thick, gray, rounded corners, fill=ForestGreen!70]
           (-1,-0.25) rectangle (-0.5,0.25);
      \draw[rounded corners,thick] (-1.0,-0.25) rectangle (1.0,0.25);
      \node {#1}; 
    }
  }
}

\node[draw, thick, shape=ellipse, minimum width = 1cm] (inp) at (2.15,0){in};

\pic at (0,0) {slimblock={GN2D}};
\node[block](sw1) at (0,-0.75){Swish};
\node[block](fir1) at (0,-1.5){FIR $\uparrow$ / $\downarrow$};
\pic at (0,-2.25) {slimblock={Conv2D 3$\times$3}};

\node[plus] (plus1) at (0,-3){$+$};

\node (emb) at (-1.75,-2) {$\mathbf{t}_{\mathrm{emb}}$};
\pic at (-1.75,-3) {slimblock={Dense}};
\node[block] (fir2) at (2.15,-3){FIR  $\uparrow$ / $\downarrow$};
\pic at (2.15,-3.75) {slimblock={Conv2D 1$\times$1}};

\pic at (0,-3.75) {slimblock={GN2D}};
\node[block] (sw2) at (0,-4.5){Swish};
\pic at (0,-5.25) {slimblock={Conv2D 3$\times$3}};

\node[plus] (plus2) at (2.15,-5.25){$+$};

\node[plus] (times) at (3.5,-5.25){$\times$};
\node (factor) at (3.5, -4.4){$1/\sqrt{2}$};

\node[draw, thick, shape=ellipse, minimum width = 1cm] (out) at (4.75,-5.25){out};

\draw[->, thick] (inp)--(1,0);
\draw[->, thick] (0,-0.25)--(sw1);
\draw[->, thick] (sw1)--(fir1);
\draw[->, thick] (fir1)--(0,-2);
\draw[->, thick] (0,-2.5)--(plus1);

\draw[->, thick] (inp)--(fir2);
\draw[->, thick] (fir2)--(2.15,-3.5);
\draw[->, thick] (2.15,-4)--(plus2);
\draw[->, thick] (plus2)--(times);
\draw[->, thick] (factor)--(times);

\draw[->, thick] (emb)--(-1.75,-2.75);
\draw[->, thick] (-0.75,-3)--(plus1);
\draw[->, thick] (plus1)--(0,-3.5);
\draw[->, thick] (0,-4)--(sw2);
\draw[->, thick] (sw2)--(0,-5);
\draw[->, thick] (1,-5.25)--(plus2);
\draw[->, thick] (times)--(out);

\node at (4,-0.75){\textbf{Slimmable Layer}};
\draw[very thick, dotted] (2.6,-0.5)rectangle (5.6,-2.5);
\pic at (4.5,-1.25) {slimblock={SlimLayer}};
\node at (4,-2){$u=0.25, 0.5,1$};
\draw[->](3.9,-1.75)--(3.8,-1.55);
\draw[->](4.5,-1.75)--(4.3,-1.55);
\draw[->](5.0,-1.75)--(5.0,-1.55);

\end{tikzpicture}
\vspace{-1em}
\caption{Residual block with slimmable channels.}
\label{fig:slimmable-block}
\vspace{-1.5em}
\end{figure}

To address this limitation, we propose SlimDiffuSE, a slimmable score model $\mathbf{F}^{u}_\phi(\mathbf{x}_t,\mathbf{y},t)$, where $u \in (0,1]$ is a continuous utilization factor controlling the active network width. 
For a given $u$, the active sub-network is obtained by applying a structured channel-wise mask that retains the first $\lceil u\,C \rceil$ channels in each slimmable layer, where $C$ denotes the full number of channels (see Fig.~\ref{fig:slimmable-block}) and $\lceil \cdot \rceil$ is the ceiling operator.
The model is trained using a multi-width optimization strategy \cite{slimtasnet} over a set of utilization factors $\mathcal{U} = \{0.25, 0.5, 1.0\}$
\begin{equation}
\vspace{-0.5em}
    \mathcal{L}(\phi) =
\mathbb{E}_{t,\mathbf{x}_0,\mathbf{y},\mathbf{z}}
\left[
\sum_{u \in \mathcal{U}} u \left\|
\mathbf{F}^{u}_\phi(\mathbf{x}_t,\mathbf{y},t)
+ \frac{\mathbf{z}}{\sigma(t)}
\right\|_2^2
\right],
\end{equation}
where $\mathbf{z} \sim \mathcal{N}_{\mathbb{C}}(\mathbf{0},\mathbf{I})$ and $\sigma(t)$ denotes the variance of the diffusion noise schedule. During inference, we perform $N$ steps in the reverse process and apply a step-dependent utilization schedule $u_n \in \mathcal{U}$, $n = 1, \dots, N$. The update rule for the reverse diffusion process is obtained as
\vspace{-0.5em}
\begin{align}
    \mathbf{x}_{n-1} = \mathbf{x}_n
    &- \gamma(\mathbf{y} - \mathbf{x}_n)\Delta t \\
    &+ g^2(t_n)\,\mathbf{F}^{u_n}_\phi(\mathbf{x}_n,\mathbf{y},t_n)\Delta t + g(t_n)\sqrt{\Delta t}\,\mathbf{z}_n,\notag
\end{align}
where $\mathbf{z}_n \sim \mathcal{N}_{\mathbb{C}}(\mathbf{0},\mathbf{I})$. This adaptive formulation enables dynamic allocation of model capacity across inference steps. Assuming the computational complexity scales linearly with the utilization factor $u$, the resulting computational cost is approximated as:
\vspace{-0.5em}
\begin{equation}
    \mathcal{C}_{\text{slim}} =2\sum_{n=1}^{N} \text{FLOPs}(\mathbf{F}^{u_n}_\phi)\approx 2\ \text{FLOPs}(\mathbf{S}_\theta)\cdot\sum_{n=1}^{N} u_n  
\vspace{-0.5em}
\end{equation}
which  can be optimized by parameterizing the slimmable score model $\mathbf{F}^{u}_\phi$, $u<1$, with lower complexity to later diffusion steps ($n$ higher), aiming at fine structure, while assigning higher model capacity for the early stages of the reverse diffusion process ($n$ small).

\noindent \textbf{SlimDiffuSE architecture:}
\label{ssub:slimmable ncsn}
In this work, we adapt the NCSN++ backbone \cite{song2020score}, a multi-resolution U-Net with skip connections and a parallel-growing U-Net path, which is used in SGMSE+~\cite{richter2023speech}. Each upsampling, downsampling, and bottleneck module is realized by residual blocks~\cite{biggan}. We use three residual blocks per upsampling module and two per downsampling module, where the last residual block handles up-/down-sampling. Each residual block consists of convolution layers, group normalization (GN), finite impulse response (FIR) filters for up/down sampling, and Swish activation, as shown in Fig.~\ref{fig:slimmable-block}. A global attention module is applied to the bottleneck at a resolution of $16 \times 16$. Fourier time embeddings $\mathbf{t}_{\mathrm{emb}}\in\mathbb{R}^{128}$ are fed to each residual block. The number of channels of the network layers is computed by multiplying the base channel number $C_{\text{base}} = 128$ with channel multipliers $(1, 1, 2, 2, 2, 2, 2)$. Different model complexities can be obtained by varying $C_{\text{base}}$. For a slimmable architecture, each residual block is converted into a slimmable residual block, where convolutions, GNs, and dense layers adapt their input/output dimensions according to a utilization factor $u \in \mathcal{U}=\{0.25, 0.5, 1\}$ following~\cite{slimtasnet}, as shown in Fig.~\ref{fig:slimmable-block}. The bottleneck attention module is likewise transformed into a slimmable attention layer controlled by $u$. This construction results in three effective model widths with $C_{\text{base}} \in \{32, 64, 128\}$.

\vspace{-1.5em}

\begin{figure}[t]
\centering
\begin{tikzpicture}[x=0.255cm, y=0.4cm, font=\scriptsize]

\definecolor{cH}{RGB}{220,80,70}    
\definecolor{cM}{RGB}{240,200,80}   
\definecolor{cL}{RGB}{120,190,120}  

\draw[<->, thick] (0,6.2) -- (30,6.2);
\node[above] at (15,6.2) {\footnotesize Total diffusion steps = 30};

\draw[fill=cH] (0,4.8) rectangle (15,5.8);
\node at (7.5,5.3) {15 (HC)};
\draw[fill=cM] (15,4.8) rectangle (25,5.8);
\node at (20,5.3) {10 (MC)};
\draw[fill=cL] (25,4.8) rectangle (30,5.8);
\node at (27.5,5.3) {5 (LC)};
\node[left] at (0,5.3) {\texttt{C1}};

\draw[fill=cH] (0,3.8) rectangle (10,4.8);
\node at (5,4.3) {10 (HC)};
\draw[fill=cM] (10,3.8) rectangle (20,4.8);
\node at (15,4.3) {10 (MC)};
\draw[fill=cL] (20,3.8) rectangle (30,4.8);
\node at (25,4.3) {10 (LC)};
\node[left] at (0,4.3) {\texttt{C2}};

\draw[fill=cM] (0,2.8) rectangle (15,3.8);
\node at (7.5,3.3) {15 (MC)};
\draw[fill=cH] (15,2.8) rectangle (25,3.8);
\node at (20,3.3) {10 (HC)};
\draw[fill=cL] (25,2.8) rectangle (30,3.8);
\node at (27.5,3.3) {5 (LC)};
\node[left] at (0,3.3) {\texttt{C3}};

\draw[fill=cM] (0,1.8) rectangle (10,2.8);
\node at (5,2.3) {10 (MC)};
\draw[fill=cH] (10,1.8) rectangle (20,2.8);
\node at (15,2.3) {10 (HC)};
\draw[fill=cL] (20,1.8) rectangle (30,2.8);
\node at (25,2.3) {10 (LC)};
\node[left] at (0,2.3) {\texttt{C4}};

\draw[fill=cL] (0,0.8) rectangle (15,1.8);
\node at (7.5,1.3) {15 (LC)};
\draw[fill=cM] (15,0.8) rectangle (25,1.8);
\node at (20,1.3) {10 (MC)};
\draw[fill=cH] (25,0.8) rectangle (30,1.8);
\node at (27.5,1.3) {5 (HC)};
\node[left] at (0,1.3) {\texttt{C5}};

\draw[fill=cL] (0,-0.2) rectangle (10,0.8);
\node at (5,0.3) {10 (LC)};
\draw[fill=cM] (10,-0.2) rectangle (20,0.8);
\node at (15,0.3) {10 (MC)};
\draw[fill=cH] (20,-0.2) rectangle (30,0.8);
\node at (25,0.3) {10 (HC)};
\node[left] at (0,0.3) {\texttt{C6}};

\draw[thick,->] (0,-1.0) -- (30,-1.0);
\node[below] at (0,-1.0) {1};
\node[below] at (30,-1.0) {0};
\node[below=2pt] at (15,-1.0) {\footnotesize Diffusion time $t$};


\end{tikzpicture}
\vspace{-1.5em}

\caption{Network configurations (\texttt{C1}--\texttt{C6}) with three model complexities assigned to different reverse diffusion steps.}
\label{fig:complexity_steps}
\vspace{-1em}
\end{figure}
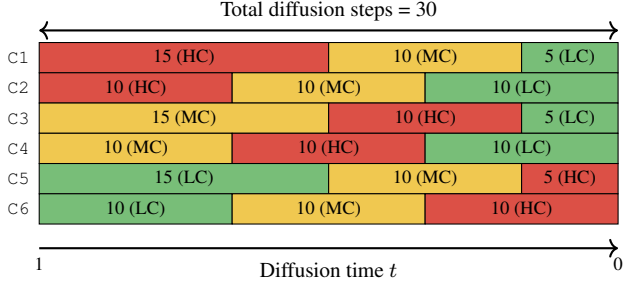

\begin{table}[t]
    \caption{Objective evaluation metrics for the three baseline complexities (HC, MC, LC) and the six predefined mixed complexity configurations C1-C6.}
    \label{tab:hypotheisresults}
    \vspace{0.5em}
    \centering
    \footnotesize 
    \setlength{\tabcolsep}{6pt} 
    \renewcommand{\arraystretch}{1.2} 
    \begin{tabular}{l 
        S[table-format=1.2(2)] 
        S[table-format=2.1(2)] 
        S[table-format=1.2(2)] 
        S[table-format=2.1(3)]} 
        \toprule
        \textbf{Model} & {\textbf{PESQ} $\uparrow$} & {\textbf{SI-SDR (dB)} $\uparrow$} & {\textbf{ESTOI} $\uparrow$} & {\textbf{SI-SIR (dB)}  $\uparrow$} \\
        \midrule
        Noisy & 1.58(46) & 9.1(55) & 0.81(12) & 9.1(5.5) \\
        \midrule
        HC    & \bfseries 2.85(72) & \bfseries 16.9(62) & \bfseries 0.93(7)  & \bfseries 27.4(9.3) \\
        MC    & 2.72(73) & 15.8(62) & 0.92(7)  & 25.9(9.7) \\
        LC    & 2.48(72) & 14.4(61) & 0.90(8)  & 23.1(9.4) \\
        \midrule
        \texttt{C1} & 2.82(70) & \bfseries 16.9(63)& \bfseries 0.93(7)  & \bfseries 27.6(9.5) \\
        \texttt{C2} & \bfseries 2.83(71) & 16.8(63) & \bfseries 0.93(7)  &  27.2(9.4) \\
        \texttt{C3} & 2.72(73) & 15.8(63) & 0.92(7)  & 26.2(10.0) \\
        \texttt{C4} & 2.72(74) & 15.9(63) & 0.92(7)  & 26.3(9.7) \\
        \texttt{C5} & 2.54(72) & 14.8(61) & 0.90(8)  & 24.4(9.3) \\
        \texttt{C6} & 2.56(71) & 14.8(60) & 0.90(8)  & 24.1(9.1) \\
        \bottomrule
    \end{tabular}
\vspace{-2em}
\end{table}

\section{Experiments and Results}
\label{sec:typestyle}
\vspace{-0.25cm}
\subsection{Experimental Details}
\noindent \textbf{Datasets:} 
Training data was generated using the Interspeech 2020 DNS Challenge dataset~\cite{reddy2020interspeech} by mixing clean speech with noise recordings at randomly sampled \acp{SNR} from $[-10, 30]\,\mathrm{dB}$ following a setup similar to~\cite{shetu2025leveraging,shetu2023ultra}. To simulate reverberation, $50\%$ of the clean utterances were convolved with randomly selected \acp{RIR} from the DNS RIR dataset~\cite{reddy2020interspeech}, which contains approximately $100$k measured and synthetic \acp{RIR} with reverberation times ranging from $0.3$ to $1.5\,\mathrm{s}$. For evaluation, the DNS Challenge non-reverberant test set~\cite{reddy2020interspeech} was used, which includes noises from $12$ VoIP-relevant categories, with \acp{SNR} in the range $[0, 25]\,\mathrm{dB}$, and consists of $150$ test samples of $10\,\mathrm{s}$ each.
\label{ssec:dataset}

\noindent \textbf{Training and evaluation:} The baseline SGMSE+ \cite{richter2023speech} and the proposed models were trained with the Adam optimizer at a learning rate of $10^{-4}$ using an STFT window size of $510$, a hop length of $128$, an FFT length of $510$, and with a sampling rate of $16$kHz. 
An exponential moving average (EMA) of the DNN weights with a decay factor of $0.999$~\cite{song2020improved} was calculated, and the best-performing model checkpoints were selected based on the highest validation \ac{PESQ} scores~\cite{rix2001perceptual}. We evaluated all methods using the following objective metrics: \ac{PESQ}~\cite{rix2001perceptual}, \ac{ESTOI}~\cite{jensen2016algorithm}, scale-invariant signal-to-distortion ratio (SI-SDR)~\cite{le2019sdr}, and scale-invariant signal-to-interference ratio (SI-SIR)~\cite{le2019sdr}. We quantified computational complexity by giga-multiply-accumulate operations per second (GMACs) and the total number of model parameters.

\vspace{-0.2cm}
\subsection{Experimental Results}
\noindent \textbf{Models at different complexity:} \label{ssub:models at diff compl} In this experiment, we show that the difficulty of the iterative denoising task in \acp{DM} for \ac{SE} varies across diffusion time steps. Exploiting this, we reduce model complexity at selected steps while maintaining performance, substantially lowering computational cost. To this end, we trained three score models based on the NCSN++ architecture: \ac{HC}, \ac{MC}, and \ac{LC}, with base channel dimensions $C_{\text{base}} \in \{128,\;64,\; 32\}$ corresponding to utilization factors  $u \in \{1.0,\;0.5,\;0.25\}$, respectively. We evaluated the assignment of these models to diffusion time steps in six predefined configurations (see Fig.~\ref{fig:complexity_steps}). The results in Tab.~\ref{tab:hypotheisresults} demonstrate that applying a low-complex model (\ac{LC}, \ac{MC}) in all reverse diffusion steps leads to performance degradation: \ac{PESQ} scores dropped by at least $0.13$ and $0.37$ for the \ac{MC} and \ac{LC} variant, respectively, compared to the \ac{HC} model. However, among the mixed-complexity configurations, variants \texttt{C1} and \texttt{C2} achieved results almost matching the performance of the full \ac{HC} evaluation. Specifically, \texttt{C1} and \texttt{C2} yielded \ac{PESQ} scores of $2.82$ and $2.83$, approaching the $2.85$ achieved by the \ac{HC} model, while reducing the overall computational load. These results also indicated that employing more model complexity at the beginning of the reverse diffusion process ($t\rightarrow1$) is important. We observed a notable performance drop for the \texttt{C3} through \texttt{C6} configurations, where the \ac{HC} model is instead employed in the middle or final segments ($t\rightarrow0$). This behavior is intuitive, as at early reverse-time steps ($t\rightarrow1$), the model must build up signal structures from pure noise, requiring higher network capacity. Conversely, as the sample approached the clean speech distribution ($t\rightarrow0$), the denoising task becomes less demanding, and a low-complexity model is sufficient for generation.

\begin{table}[t]
    \caption{Objective evaluation metrics for slimmable models and their mixed-complexity configurations.}
    \label{tab:merged_slimmable}
    \vspace{0.5em}
    \centering
    \footnotesize
    \setlength{\tabcolsep}{4pt}
    \renewcommand{\arraystretch}{1.2}
    \begin{tabular}{l c c c c}
        \toprule
        \textbf{Model} & {\textbf{PESQ} $\uparrow$} & {\textbf{SI-SDR (dB)} $\uparrow$} & {\textbf{ESTOI} $\uparrow$} & {\textbf{SI-SIR (dB)}  $\uparrow$} \\
        \midrule
        $u=1.0$   & \bfseries  2.78 $\pm$ 0.68 & \bfseries  16.8 $\pm$ 5.7 & \bfseries  0.92 $\pm$ 0.07 & \bfseries  28.9 $\pm$  9.2 \\
        $u=0.5$   & 2.57 $\pm$ 0.75 & 15.6 $\pm$ 6.0 & 0.90 $\pm$ 0.08 & 26.8 $\pm$ 9.3 \\
        $u=0.25$  & 2.18 $\pm$ 0.70 & 13.3 $\pm$ 5.8 & 0.88 $\pm$ 0.10 & 23.1 $\pm$ 10.0 \\
        \midrule
        \texttt{C1} & 2.76 $\pm$ 0.67 & 16.7 $\pm$ 5.6 & \bfseries 0.92 $\pm$ 0.07 & 29.2 $\pm$ 8.8 \\
        \texttt{C2} & \bfseries  2.77 $\pm$ 0.67 & \bfseries  16.8 $\pm$ 5.6 & \bfseries  0.92 $\pm$ 0.07 & \bfseries  29.5 $\pm$ 9.1 \\
        \texttt{C3} & 2.56 $\pm$ 0.74 & 15.6 $\pm$ 5.9 & 0.90 $\pm$ 0.08 & 26.8 $\pm$ 8.9 \\
        \texttt{C4} & 2.60 $\pm$ 0.74 & 15.8 $\pm$ 6.0 & 0.91 $\pm$ 0.08 & 27.1 $\pm$ 9.0 \\
        \texttt{C5} & 2.28 $\pm$ 0.71 & 14.0 $\pm$ 5.7 & 0.88 $\pm$ 0.10 & 25.1 $\pm$ 9.1 \\
        \texttt{C6} & 2.28 $\pm$ 0.70 & 14.0 $\pm$ 5.7 & 0.88 $\pm$ 0.09 & 25.3 $\pm$ 9.7 \\
        \bottomrule
    \end{tabular}
\vspace{-2em}
\end{table}

\begin{table}[t]
    \caption{Objective performance across search-optimized schedules. Intervals $[s_{\text{start}}, s_{\text{end}}]$ denote reverse sampling steps where specific utilization factors $u$ are active.}
    \vspace{0.5em}
    \label{tab:optimized_schedules}
    \centering
    \footnotesize
    \setlength{\tabcolsep}{5pt}
    \begin{tabular}{c c c c c c c}
    \toprule
    \multicolumn{3}{c}{\textbf{Complexity Transitions }} & 
\makecell{\textbf{Avg.}\\\textbf{$u$}} & 
\makecell{\textbf{GMACs}\\\textbf{/Frame}} & 
\makecell{\textbf{PESQ}\\$\uparrow$} & 
\makecell{\textbf{SI-SDR}\\ (dB) $\uparrow$} \\
\cmidrule(r){1-3}
$u=1.0$ & $u=0.5$ & $u=0.25$ &  &  &  &  \\
    \midrule
    $[1, 1]$   & $[2, 6]$   & $[7, 30]$  &  \bfseries  0.317 & \bfseries  15.62 & 2.77 &  16.7 \\
    $[1, 2]$   & $[3, 7]$   & $[8, 30]$  & 0.342 & 19.54 & 2.76 & 16.6 \\
    $[1, 3]$   & $[4, 8]$   & $[9, 30]$  & 0.367 & 23.46 & \bfseries  2.78 & 16.7 \\
    \midrule
    $[1, 5]$   & $[6, 10]$  & $[11, 30]$ & 0.417 & 31.30 & 2.76 & 16.6 \\
    $[1, 7]$   & $[8, 12]$  & $[13, 30]$ & 0.467 & 39.14 & 2.76 & 16.6 \\
    $[1, 10]$  & $[11, 20]$ & $[21, 30]$ & 0.542 & 50.90 & 2.77 & \bfseries 16.8 \\
    $[1, 15]$  & $[16, 25]$ & $[26, 30]$ & 0.667 & 70.50 & 2.76 & 16.7 \\
    \midrule
    \ac{HC} & --- & --- & 1.000 & 125.40 & 2.85 & 16.9\\
    $u=1.0$ & --- & --- & 1.000 & 125.40 & 2.78& 16.8\\  
    \bottomrule
    \end{tabular}
\vspace{-2.2em}
\end{table}
\noindent \textbf{Slimmable diffusion model:}
While our initial experiments validated that predefined combinations of models of varying width could match the performance of the full \ac{HC} model with significantly reduced computational complexity, this approach increased the overall memory footprint. For example, the \texttt{C2} configuration (see Tab.~\ref{tab:hypotheisresults}) reduced complexity by over $56\%$ (from $125$ to $54.8$ GMACs/Frame), but increased the total number of parameters from $65$ M to $89$ M as three separate score models with $65$, $18.1$, and $5.5$M parameters have to be maintained. In contrast, as outlined in Sec.~\ref{sec:format}, a slimmable model with variable width adapts its complexity across diffusion steps without incurring the parameter overhead of multiple models. Tab.~\ref{tab:merged_slimmable} presents the results for SlimDiffuSE across different utilization factors $u$ for $30$ reverse-diffusion steps. The full-width configuration ($u=1.0$) achieved performance comparable to the \ac{HC} baseline. However, reduced-width configurations ($u=0.5, 0.25$) show some degradation compared to the standalone \ac{MC} and \ac{LC} models (see Tab.~\ref{tab:hypotheisresults}), since parameter sharing across widths imposes a joint optimization objective on the shared weights. In particular, \ac{PESQ} drops to $2.57$ and $2.18$, compared to $2.72$ and $2.48$ for \ac{MC} and \ac{LC}, respectively. Despite this, when using predefined mixed-complexity configurations (replacing \ac{HC}, \ac{MC}, and \ac{LC} with corresponding SlimDiffuSE configurations), configurations \texttt{C1} and \texttt{C2} achieved performance comparable to the original mixed-complexity results reported in Tab.~\ref{tab:hypotheisresults} for these configurations. 

\begin{algorithm}[ht]
\small
    \caption{Greedy search over utilization factors.}
    \vspace{0.5em}
    \label{alg:greedy_search}
    \begin{algorithmic}
    \STATE \textbf{Input:} Number of diffusion steps $T$, utilization factors $\mathcal{U} = \{1.0, 0.5, 0.25\}$, tolerance $\epsilon=0.1$, validation set $\mathcal{D}_{\mathrm{val}}$
    \STATE \textbf{Initialize:} schedule $S = [1.0, \dots, 1.0]$
    \STATE \textbf{Initialize:} set of valid configurations $\mathcal{S}_{\mathrm{all}} = \{S\}$
    \STATE $P_{\mathrm{base}} \leftarrow \mathrm{PESQ}(S, \mathcal{D}_{\mathrm{val}})$
    
    \FOR{$u = 0.5, 0.25$}
        \STATE $i \leftarrow 1$
        \WHILE{$i \leq T$}
            \STATE $S' \leftarrow S$ 
            \STATE $S'_{T-i+1:T} = u$
            \STATE \textbf{if} $\mathrm{PESQ}(S', \mathcal{D}_{\mathrm{val}}) < P_{\mathrm{base}} - \epsilon$ \textbf{then break}
            \STATE $S \leftarrow S'$, $i \leftarrow i + 1$
            \STATE $\mathcal{S}_{\mathrm{all}}\leftarrow \{S\} \cup \mathcal{S}_{\mathrm{all}}$
        \ENDWHILE
    \ENDFOR
    
    \STATE \textbf{return} $\mathcal{S}_{\mathrm{all}}$
   
    \end{algorithmic}
    
\end{algorithm}

\noindent \textbf{Greedy search:}
We further aim to determine an optimal allocation of model widths along the reverse-diffusion steps for SlimDiffuSE. To this end, we adopted an evolutionary-inspired greedy search algorithm that operated over the set of utilization factors $\mathcal{U} = \{1.0, 0.5, 0.25\}$ as shown in Alg.~\ref{alg:greedy_search}. Starting from a fully utilized model ($u=1.0$), the method progressively replaced later diffusion steps with lower utilization factors, while maintaining performance within a tolerance $\epsilon$ relative to the baseline  ($u=1.0$). In this work, the tolerance $\epsilon$ is based on the objective metric PESQ, calculated on a validation dataset $\mathcal{D}_{\mathrm{val}}$ (a smaller disjoint subset of the training dataset) for a given reverse diffusion schedule $S$. We enforced a monotonic, non-increasing utilization constraint for decreasing $t$, which significantly reduced the search space. Thus, instead of evaluating all $L^T$ configurations ($L=|\mathcal{U}|$), the problem reduced to identifying the transition boundaries, resulting in $\mathcal{O}(LT)$ complexity, instead of  $\mathcal{O}(L^T)$. For $L=3$ and $T=30$, this reduced evaluations from $3^{30} \approx 2.05 \times 10^{14}$ to at most $60$.
As summarized in Tab.~\ref{tab:optimized_schedules}, the proposed greedy search strategy identified highly efficient configurations that maintain competitive performance while drastically reducing computational overhead. In particular, an optimized configuration with average utilization of $0.317$ and $15.62$ GMACs/Frame achieved a \ac{PESQ} of $2.77$, comparable to the HC baseline ($2.85$) and SlimDiffuSE full-width variant ($2.78$ for $u=1.0$). Overall, this corresponded to an $87.5\%$ reduction in computational cost relative to the HC model, demonstrating that competitive performance can be achieved even when most of the denoising trajectory is offloaded to lower-complex sub-networks.
\vspace{-1.2em}

\section{Conclusions}
\label{sec:majhead}
\vspace{-1em}
In this work, we showed that the difficulty of the per-step denoising task in diffusion-based \ac{SE} varies along the diffusion trajectory. Our proposed SlimDiffuSE uses a greedy search algorithm for optimizing the slimming schedule and achieves performance comparable to a full high-complexity model while significantly reducing computational cost. We believe our work establishes a flexible trade-off between \ac{SE} quality and efficiency, paving the way for real-time diffusion-based enhancement on resource-constrained hardware.


\let\oldbibliography\thebibliography
\renewcommand{\thebibliography}[1]{%
  \oldbibliography{#1}%
  \footnotesize 
  \setlength{\itemsep}{-0.0ex} 
  \setlength{\parsep}{0pt}  
  \setlength{\parskip}{0pt} 
  \setlength{\leftmargin}{1em} 
}

\bibliographystyle{IEEEbib}
\bibliography{strings,refs}
\end{document}